\documentclass[aps,amsmath,amssymb,showkeys,twocolumn]{revtex4-1}
\usepackage{epsfig}
\graphicspath{{./figures/}}
\usepackage{color}
\begin{document}

\title{Role of helicity in DNA hairpin folding dynamics}

\author{Huaping Li} \affiliation{Department of Physics, Ko\c c University, Istanbul, 34450, Turkey}

\author{Alkan Kabak\c c\i o\u glu\thanks{akabakcioglu@ku.edu.tr}}
\email[]{akabakcioglu@ku.edu.tr}
\affiliation{Department of Physics, Ko\c c University, Istanbul, 34450, Turkey}

\begin{abstract}
  We study hairpin folding dynamics by means of extensive computer
  simulations, with particular attention paid to the influence of
  helicity on the folding time $\tau$. We find that the dynamical
  exponent $\alpha$ of the anomalous scaling $\tau \sim N^\alpha$ for
  a hairpin with length $N$ changes from $1.6$ ($\simeq 1+\nu$) to
  $1.2$ ($\simeq 2\nu$) in three dimensions, when duplex helicity is
  removed. The relation $\alpha=2\nu$ in rotationless hairpin folding
  is further verified in two dimensions ($\nu=0.75$), and for a
  ghost-chain ($\nu=0.5$). This, to our knowledge, is the first
  observation of the theoretical lower bound on $\alpha$, which was
  predicted earlier on the basis of energy conservation for polymer
  translocation through a pore. Our findings suggest that the folding
  dynamics in long helical chains is governed by the duplex dynamics,
  contrasting the earlier understanding based on the stem-flower
  picture of unpaired segments. We propose a scaling argument for
  $\alpha=1+\nu$ in helical chains, assuming that duplex relaxation
  required for orientational positioning of the next pair of bases is
  the rate-limiting process.

\end{abstract}

\keywords{hairpin folding, anomalous dynamics, helicity, rotational
  relaxation}

\maketitle

\newpage

DNA/RNA hairpin folding is the temperature-driven self assembly of a
palindromic nucleic acid composed of two complementary sequences
linked by a relatively short ``loop'' segment. Transcription and
folding of small hairpins (e.g., siRNAs, miRNAs) help initiate
biochemical reactions, cell signaling, gene expression, viral
response, in many organisms~\cite{Zhuang2000, Noller2005, Pan2006,
  Solomatin2010}. Their synthetic counterparts are used ubiquitously
in biotechnological applications, such as in CRISPR ~\cite{Peters2015,
  Teotia2016, Kim2017, Dowdy2017}.  Interest in the folding dynamics
of such molecules has grown recently, due to the availability of new
experimental techniques which allow high-resolution observations both
in time and space~\cite{Rouskin2014, Solem2015, Ritchie2015}. As
revealed by numerical simulations and experiments, formation of a
folding nucleus at the center of the hairpin is the time-limiting
step~\cite{Zhang2002, Jung2006, Ma2007, Jung2008,
  Ansari2001}. Nonetheless, actual folding time (zippering after
nucleus formation, sometimes referred as the transition-path time) has
been the focus of several recent studies due to its anomalous
character~\cite{Ferrantini2011, Sakaue2017, Frederickx2014}.

Progression of zippering can be monitored through the duplex length,
$n(t)$, which serves as the natural reaction coordinate. Earliest
theoretical models for predicting the folding time, such as the zipper
model, were based on the equilibrium free-energy difference between
paired (double-strand) and unpaired (single-strand) states. Such
considerations predict a ballistic process $n(t)\sim t$ for $T<T_c$
and a diffusive one for $T=T_c$, where $T_c$ is the folding
temperature~\cite{Cocco2003, Richard2003, Poland1966, Fisher1966,
  Sakaue2017}. Yet, experimental data appear to yield a better fit to
the scaling relation $n(t)\sim t^{1/\alpha}$ with $\alpha =
(1+\nu)$~\cite{Neupane2012, Frederickx2014}.

The mechanism for the observed anomalous dynamics has been
investigated by exploiting the analogy to field-driven polymer
translocation across a membrane (Y-junction of the folding hairpin
corresponding to the membrane pore)~\cite{Ferrantini2011, Manghi2016,
  Vocks2008, Palyulin2014}. In fact, the exponent $(1+\nu)$ has been
previously reported for the translocation-time {\em vs.}  polymer
length~\cite{Kantor2004}, under the assumption that the polymers on
both sides of the pore are in quasi-equilibrium at all times. Yet, as
several studies pointed out~\cite{Ferrantini2011, Frederickx2014},
hairpin folding is an out-of-equilibrium phenomenon, therefore,
observation of identical exponents in the two processes is conceivably
coincidental. It was recently argued that the anomalous scaling of the
hairpin folding time follows from Langevin dynamics under constant
force, with a friction term associated with the relatively stretched
portions of the unfolded arms~\cite{Frederickx2014}. While the
time-limiting process in the folding dynamics is still unclear
(further discussed below), numerical value of $\alpha$ is also subject
to continuing debate since out-of-equilibrium translocation processes
and Monte-Carlo simulations of hairpin folding on lattice models also
show a regime with $\alpha =
(1+2\nu)/(1+\nu)$~\cite{Ferrantini2011,Palyulin2014,Luo2009}.

A marked difference between hairpin folding and polymer translocation
phenomena is the rotational aspect of the dynamics, in the former
case, induced by the natural twist of the DNA/RNA duplex. Despite past
and recently renewed interest in statistical and dynamical properties
of (un)winding polymers~\cite{Baiesi2010, Walter2014}, existing
studies on hairpin folding pay no attention to implications of duplex
helicity. We here address the role of twist on zippering dynamics by
comparing the folding rates of two computational models which are
almost identical, except for the difference in angle/dihedral
potentials which induces an inherent twist in one model (helix) and
but not in the other (ladder). By performing molecular dynamics (MD)
simulations on chains more than an order of magnitude longer than the
persistence length of the duplex, we demonstrate that twist is, in
fact, an essential factor in determining the folding-time scaling. As
a bonus, the ladder model emerges, to our knowledge, as a unique
example of hairpin folding which realizes the lower bound
$\alpha=2\nu$ imposed by energy conservation~\cite{Vocks2008}.

\begin{figure}
  \vspace*{12pt}
  \includegraphics[width=0.55\linewidth]{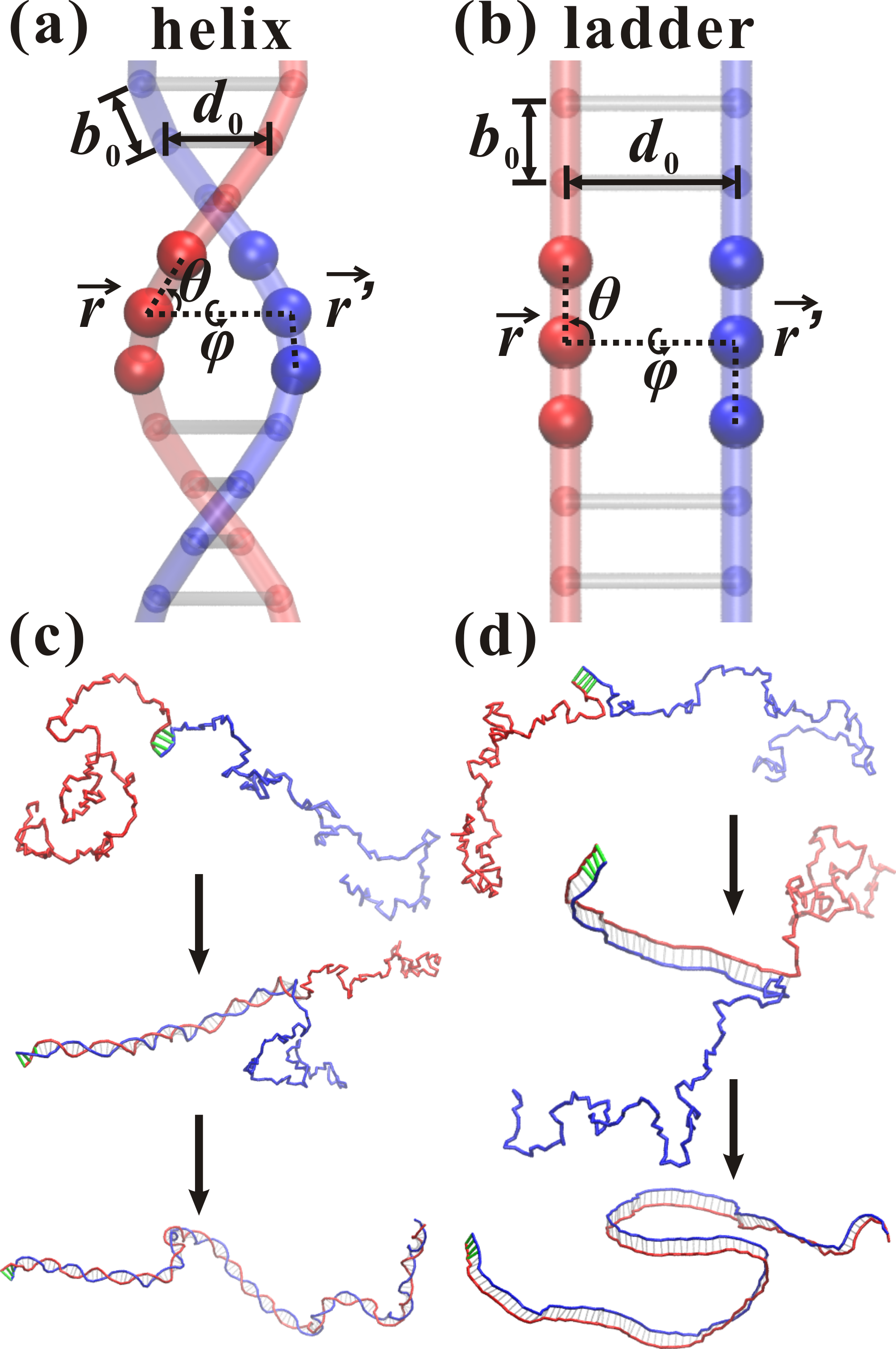}
  \caption{\label{fig_model} Schematic view of the coarse-grained helix
  and ladder models (a, b) and their folding processes (c, d). $\theta$
  and $\varphi$ denote the inter-chain angle and dihedral potentials.
  $b_0$ and $d_0$ are the equilibrium bond lengths. $\vec{r}$ and $\vec{r'}$
  are the radius vectors for the complementary bases.}
\end{figure}

We use a coarse-grained one-bead-per-base model (see
Fig.~\ref{fig_model}a and Fig.~\ref{fig_model}b) where a single DNA
strand is held together by harmonic bonds with an equilibrium length
$b_0$ and rotationally free joints at bead positions. Hairpin duplex
is modeled to be composed of complementary bases occupying symmetric
positions relative to the center. Base-pairing (inter-strand)
interaction is a segmented potential which has a minimum value of
$\epsilon_{bp}$ at pair distance $d_0$, vanishes beyond a maximum bond
distance $d_{max}$, and is specific (each base is allowed to bond with
its complement only). Pairing also induces inter-strand angle and
dihedral potentials which yield a DNA-like structure in the helical
model in Fig.~\ref{fig_model}a and a zero-twist structure in the
``ladder'' model in Fig.~\ref{fig_model}b. The local and non-local
(excluded volume) potentials are used to enforce self-avoidance except
for the ``ghost'' chain simulations. Associated potential functions
and parameters are given in the supporting information.  Hydrodynamic
and electrostatic interactions are not included, hence Rouse dynamics
is applicable.

MD trajectories are obtained by means of Langevin dynamics in NVT
ensemble, implemented in C{\tiny ++} for speed. Prior to folding
simulations, the critical temperature $T_c$ was obtained separately
for ladder and helix models by setting $\lambda(T_c)=0.5$, where
$\lambda$ is the mean pair fraction. Note that, contrary to one's
``mechanical'' intuition, the helical structure folds somewhat easier
than the ladder ($T_c^{l}/T_c^{h} \simeq 1.08$), as a result of the
smaller entropy of the helical duplex. The mismatch in duplex
entropies is due to the difference in persistence lengths (34.3 bps
and 23.3 bps for the helical and ladder models, respectively), a
consequence of twist-bend coupling~\cite{Marko1994}. Note that, even
the helical model is not a faithful representation of the actual DNA
structure, but it captures the essential physical ingredients for the
subject of this study and is simple enough to study long
chains. Temperature in all of our MD simulations was chosen to be
$0.16$ in units of the pair bonding energy $\epsilon_{bp}$. Time is
given in dimensionless units in all figures.

In order to investigate folding dynamics (see Fig.~\ref{fig_model}c
and Fig.~\ref{fig_model}d for snapshots) we followed the procedure
outlined in Ref.~\cite{Frederickx2014}. In particular, the nucleation
stage was bypassed by starting the MD simulations from an unfolded
chain which is equilibrated {\em a priori} at $T > T_c$ and has its
first four base pairs (at the center of the polymer) permanently
bound. The end effects were removed by defining the folding time
$\tau$ as $n(\tau)/N = 0.75$. In order to elucidate the scaling
behavior $\tau\sim N^\alpha$, we covered a wide range of hairpin
lengths in the interval $24\le N\le 644$. Note that, lattice models
can probe even larger systems sizes, but they do not faithfully
represent the helical duplex structure and associated rotational
dynamics that we underline below.

\begin{figure}
  \vspace*{12pt}
  \includegraphics[width=0.75\linewidth]{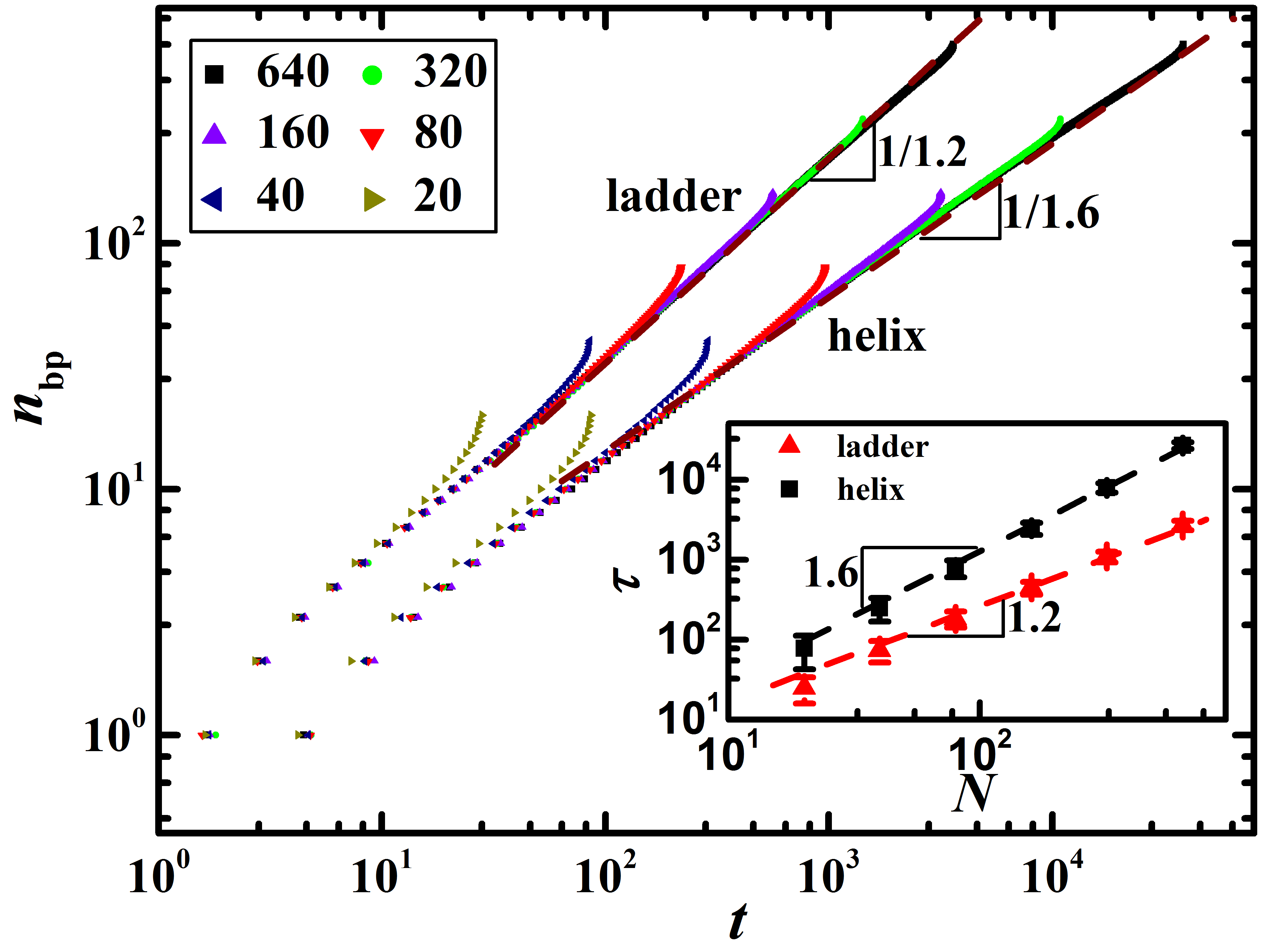}
  \caption{\label{fig_3d} Number of the formed base pairs as a
    function of time for the helix and ladder structures in three
    dimensions. The wine color dashed lines correspond to slopes of $1/1.6$
    and $1/1.2$ for the helix and ladder, respectively.  The inset
    shows the total folding time {\em vs.} the length $N$ of the
    hairpin.}
\end{figure}

Our central result is given in Fig.~\ref{fig_3d} where we plot both $n(t)$
{\em vs.} $t$ and the mean folding time $\tau$ {\em vs.} $N$ (all
averaged over $\sim 10^3$ independent runs), separately for the
helical and the ladder models. Data points spanning more than a decade
in $N$ are consistent with $\tau \sim N^{1+\nu}$ for the helical
model. This exponent was reported earlier in Ref.~\cite{Frederickx2014} where
significantly shorter chains were investigated by means of a
3-beads-per-nucleotide hairpin model~\cite{Sambriski2009}. The ladder model
obeys a visibly different scaling law which we postulate to be $\tau
\sim N^{2\nu}$ (also shown in Fig.~\ref{fig_3d} for comparison).

The difference between the helix and ladder folding times stands in
contrast with the heuristic understanding of $\alpha=(1+\nu)$
developed in Ref.\cite{Frederickx2014}, where the time evolution of
the duplex length $n(t)$ was proposed to obey the Langevin equation
$\gamma(n)\dot{n} = f $, with $f$ a constant binding force satisfying
$fa/k_BT \gtrsim 1$ and $\gamma(n) \sim n^{\alpha-1}$ representing the
friction on the stretched ``stems'' of the unfolded segments which are
being pulled towards the Y-junction. In this scenario, the folding
speed is determined solely by the unpaired, single-strand portion of
the hairpin. Accordingly, $\alpha$ is expected to remain unchanged
after the helical duplex is replaced by a non-helical ladder
geometry. The contrast between this expectation and our numerical
results in Fig.~\ref{fig_3d} is striking, especially in view of the
observation of a streched region in both systems (not shown here). The new exponent
$\alpha \simeq 1.2$ we find for ``ladder folding'' is also visibly
different from $\alpha = (1+2\nu)/(1+\nu) \simeq 1.37$ found in another
out-of-equilibrium stress propagation model for DNA translocation through
a pore~\cite{Vocks2008}.

\begin{figure}
  \vspace*{12pt}
  \includegraphics[width=0.75\linewidth]{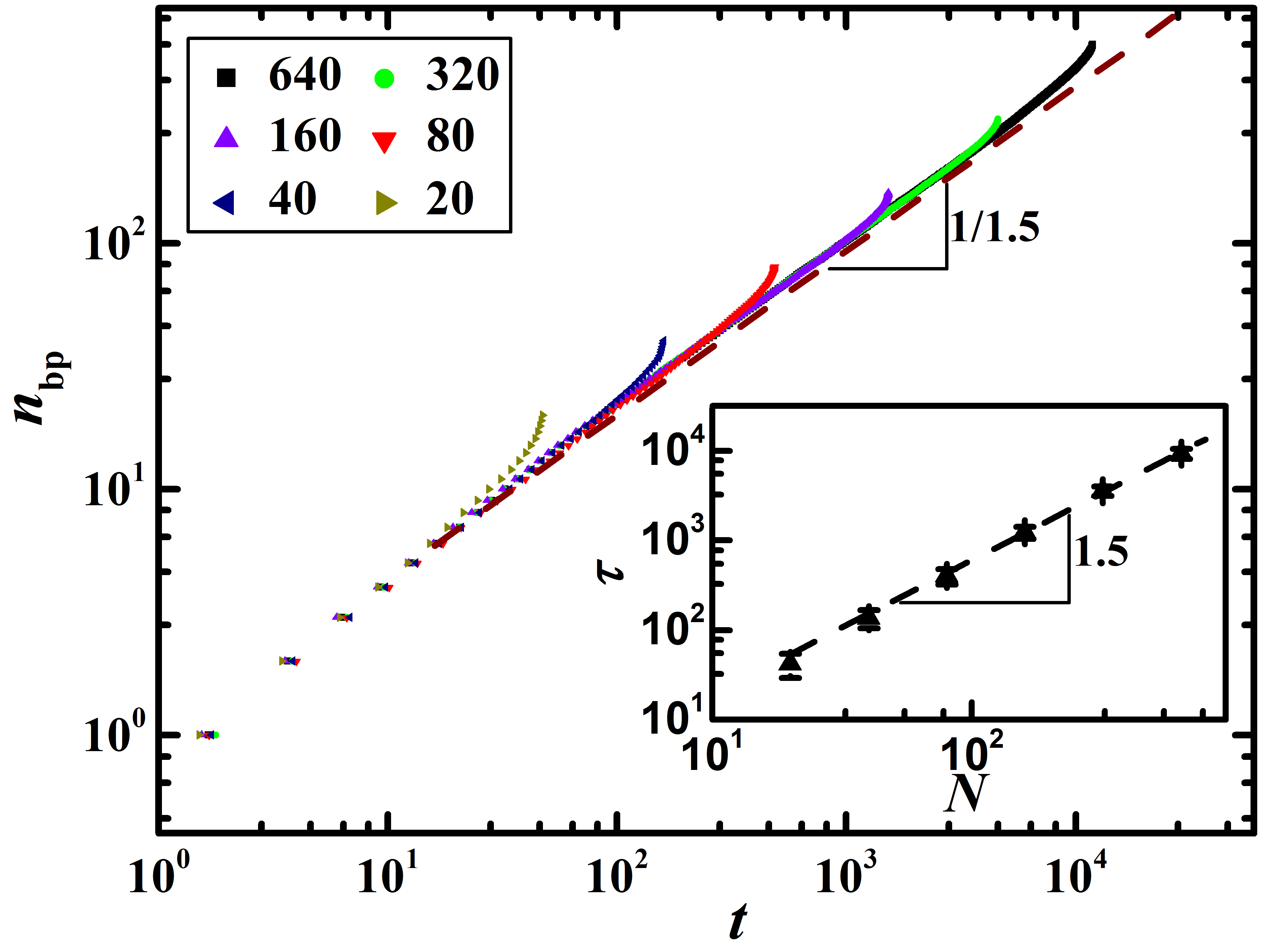}
  \caption{\label{fig_ladder_2d} Number of the formed base pairs as a
    function of time for the ladder in two dimensions. The wine color
    dashed line shows the slope of $1/1.5$.  The inset shows the total
    folding time {\em vs.} the length $N$ of the hairpin.}
\end{figure}

We propose that $\alpha=2\nu$ for the ladder model and put it to test
in two alternative settings where the Flory exponent is modified by
(a) changing the dimension, (b) removing self-avoidance. As for (a),
the obvious choice is to confine the ladder-like hairpin to two
dimensions, since its equations of motion can trivially be constrained
to a plane (in contrast, the finite thickness of the helical duplex
makes confinement of the single-strand portions technically
difficult). This is a convenient test ground where $(1+\nu) = 1.75$
and $2\nu = 1.5$ are also easy to distinguish. The analysis of our MD
simulations reported in Fig.~\ref{fig_ladder_2d} are in excellent
agreement with $\alpha = 2\nu$.

\begin{figure}
  \vspace*{12pt}
  \includegraphics[width=0.75\linewidth]{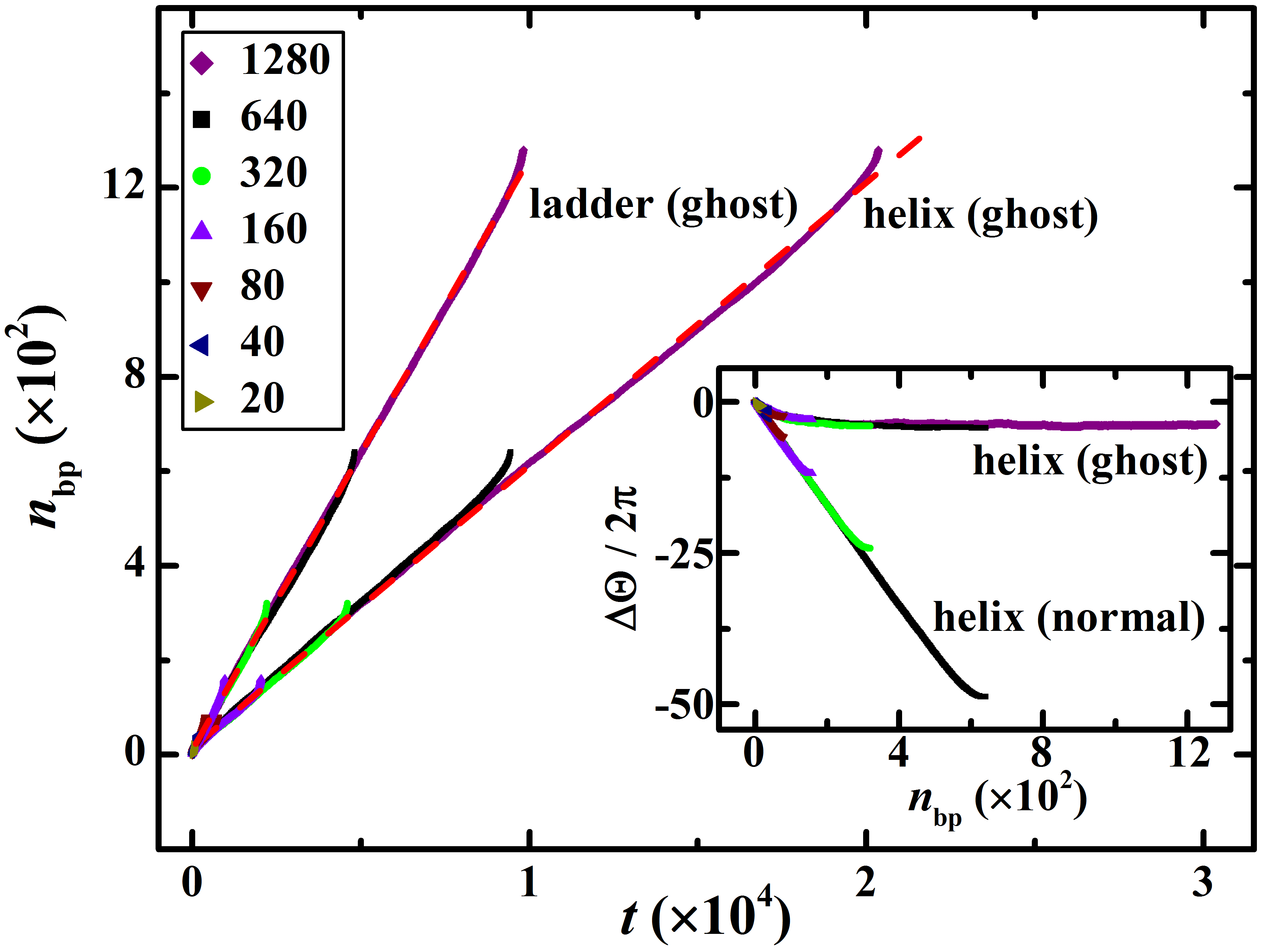}
  \caption{\label{fig_ghost} Number of the formed base pairs as a
    function of time for the helix and ladder structures using a
    ``ghost'' model in three dimensions. The red color dashed lines
    correspond to constant folding rates. The inset shows the
    comparison of the rotation angle ($\Delta\Theta$) of duplex (about the centerline)
    in the ``ghost'' and normal (self-avoiding) helical models.}
\end{figure}

Scenario (b) was implemented by removing the hard-core repulsion term
in the model, hence producing a ``ghost'' polymer with
$\nu=0.5$. Interestingly, corresponding folding dynamics (shown in
Fig.~\ref{fig_ghost}) is now ballistic {\em both for the ladder-like
  and the helical} hairpin models, where the duplex length $n(t)$ is
proportional to the elapsed time with model-dependent growth
rates. While ladder folding still conforms with $\alpha=2\nu$, hence
providing further support for our hypothesis, observation of the same
exponent in the helical case disagrees with the above picture. Next,
we address this issue.

An obvious difference between the dynamics of the two models is the
rotational aspect of the folding process in the helical model. Since
the single-stranded portion is much harder to rotate (except for the
very last stages of folding), the helical duplex has to rotate around
the centerline as it folds. Ladder model is not subject to such a
constraint. We therefore check if the ``ghost'' helical model above is
any different in this respect. The inset of Fig.~\ref{fig_ghost} shows
the rotation angle of the duplex around its centerline as a function
of time for the original and the ``ghost'' helical models. In fact, in
absence of self-avoidence, the helical duplex folds practically
without any rotation, i.e., the ``ghost'' helical model is
rotationally more similar to the ladder model. Hence, we conclude that
the slow ($\alpha=1+\nu$) and fast ($\alpha=2\nu$) folding behaviors
observed above are linked to duplex rotation.

It is worthwhile to point out that $\alpha=2\nu$ is a lower bound set
by the folding energetics. To show this, we adopt an argument from
polymer translocation studies~\cite{Vocks2008} to the present context:
Given $\tau(n)\sim n^\alpha$, the mean velocity of the bases with
index $n$ (between the beginning of the folding process and the
instant they pair up) is
\begin{equation}
v(n) \sim \frac{|\vec{r}(n)-\vec{r'}(n)|}{n^\alpha} \sim n^{\nu-\alpha}
\end{equation}
where $\vec{r}$ and $\vec{r'}$ are the radius vectors for the
complementary bases. The energy lost to friction can then be expressed
as
\begin{equation}
E_f \sim \int_0^N \eta\, v(n)\,n^\nu\,dn \sim \eta N^{2\nu-\alpha+1}
\end{equation}
where $\eta$ is the friction coefficient. Work done against friction
is provided by base-pair binding during folding, therefore we expect
\begin{equation}
N^{2\nu-\alpha+1} \le N \ \ \mbox{or} \ \ \ \alpha \ge 2\nu\ .
\end{equation}

\begin{figure}
  \vspace*{12pt}
  \includegraphics[width=0.75\linewidth]{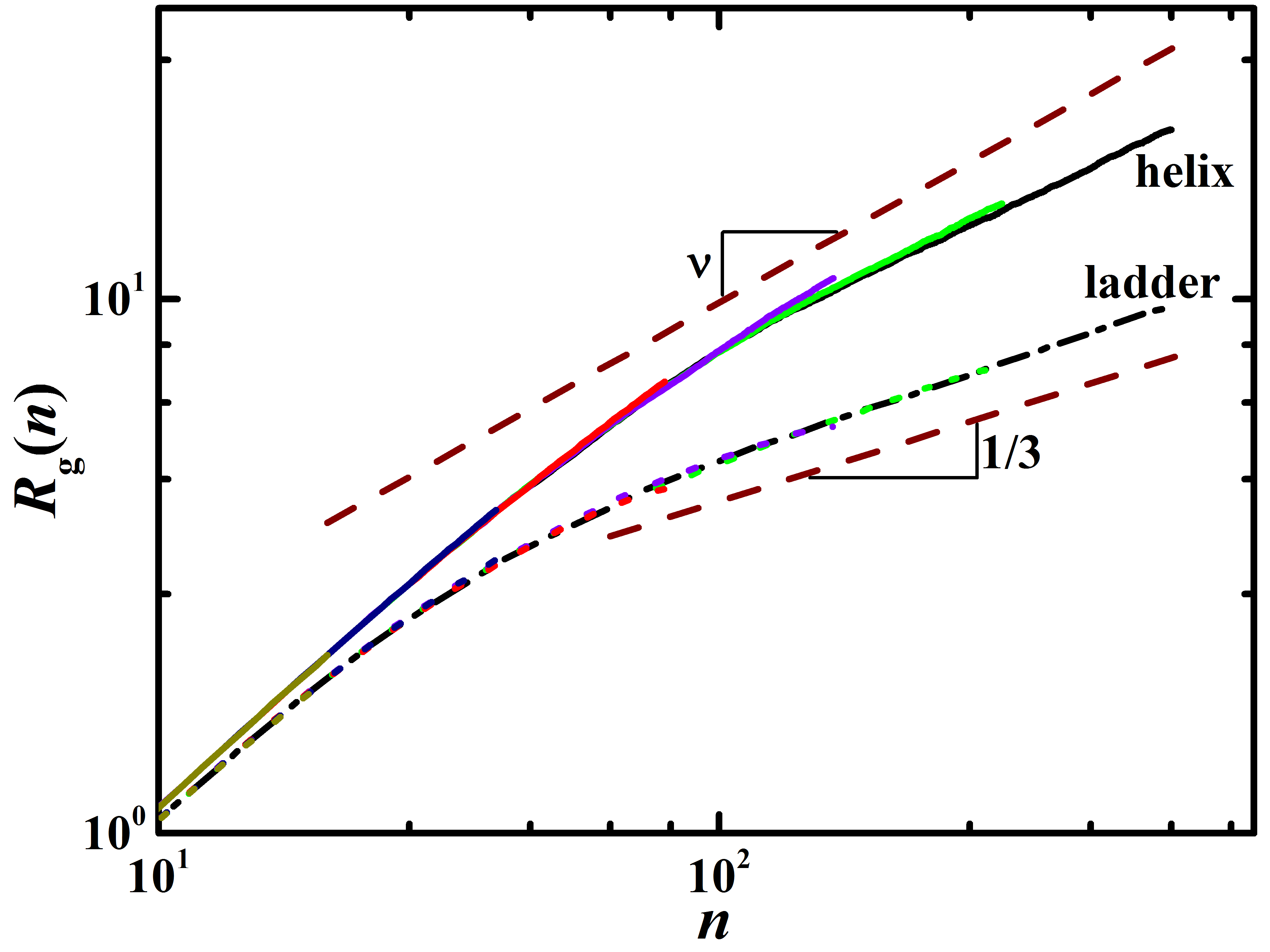}
  \caption{\label{fig_Rg} Radius of gyration of the duplex during the
  folding processes of the helix and ladder. The wine color dashed lines
  indicate the upper bound ($\nu$) and lower bound ($1/3$) for the slopes
  corresponding to equilibrated and fully compact structures, respectively.}
\end{figure}

Our observations suggest that hairpin folding is a complex
non-equilibrium phenomenon involving competing rotational and
translational processes. In absence of helicity, folding progresses at
a speed limited only by the constraint on the rate of energy transfer
between bonds forming at the Y-junction and the viscous
environment. On the other hand, folding rate of an DNA-like hairpin
structure is significantly slower due to the required rotational
relaxation of the duplex (as implied by Fig.~\ref{fig_ghost} inset).

Note that, a polymer translocating through a pore also displays a fast
and a slow regime, depending on the applied force. Translocation time
({\em vs.} length) exponents $\alpha = (1+2\nu)/(1+\nu)$ (attributed to
stress propagation dynamics) for fast translocation and
$\alpha=(1+\nu)$ for slow translocation~\cite{Luo2009} are analogous
to the hairpin folding scenario here. In Fig.~\ref{fig_Rg} we show
that, in the slow-folding regime the helical duplex is not far from
equilibrium. On the other hand, the fast-folding ladder structure
maintains a duplex which is very compact (more so than the trans
portion in fast translocation simulations~\cite{Luo2009}).

Motivated by these observations, below we propose a scaling argument
for $\alpha=(1+\nu)$ in helical hairpin folding. Given that the
relaxation of the duplex is the rate-limiting step, assume that the
duplex with length $n(t)$ has size $\sim n(t)^\nu$ at all times. A
Langevin equation for the folding process can then be written as
\begin{equation}
	\eta(n)dx/dt \sim f,
\label{eq4}  
\end{equation}
\noindent
where $\eta(n) \sim n$ is the friction coefficient, $dx$ is the
displacement of the duplex due to an added pair:
\begin{equation}
	dx = (n+dn)^\nu -n^\nu \sim \nu n^{\nu-1}dn
\end{equation}
\noindent
and $f$ is the force at the Y-junction due to the binding potential,
which can be considered constant. Then Eq.~(\ref{eq4}) becomes
\begin{equation}
	n \cdot n^{\nu-1}dn \sim dt.
\label{eq6}  
\end{equation}
Integrating Eq.~(\ref{eq6}) yields the folding time, $\tau \sim
N^{1+\nu}$. Note that, unlike the polymer translocation problem, the
mechanical properties of the polymer on the two sides of the
Y-junction are quite different. This introduces different relaxation
time scales to the picture, allowing one side (duplex) to maintain a
quasi-equilibrium state throughout most of the folding, while the
other (unpaired) segments are out of equilbrium. In the ladder
scenario, the rotational relaxation of the duplex is not a
prerequisite for pairing, hence folding progresses with both sides
visibly out of equilibrium. In this case, we observed that the folding
rate is limited only by the allowed rate of energy flow to the fluid,
given the initial equilibrium configuration of the hairpin. A thorough
investigation of this regime is needed to formulate a more
satisfactory theory of this fast folding regime. In conclusion, our
findings provide a novel perspective on hairpin folding dynamics by
unveiling the significant, and so far ignored, contribution of the
rotational motion of the duplex in the process.

\subsection{Acknowledgements}
We are thankful to E. Carlon for his helpful comments on our
preliminary results. We are also in debt with M. \" Ozt\" urk and
M. Sayar for their contributions during model development. This work
is supported by TUBITAK through the grant MFAG-114F348.


\bibliographystyle{apsrev4-1}
\bibliography{references}
\end{document}